\documentclass[aps,prd,twocolumn,showpacs,amsmath,amssymb]{revtex4-1}
\usepackage{amsmath}
\usepackage{graphicx}
\usepackage{subfigure}
\usepackage{epstopdf}
\usepackage{color}
\usepackage{multirow}
\usepackage{setspace}
\usepackage{overpic}
\usepackage{amssymb}
\usepackage[colorlinks,urlcolor=blue, citecolor=blue,linkcolor=blue] {hyperref}
\usepackage{lineno}
\usepackage{bm}

\usepackage{rotating}
\usepackage[utf8]{inputenc}
\let\oldequation\equation
\let\oldendequation\endequation
\renewenvironment{equation}
 {\linenomathNonumbers\oldequation}
 {\oldendequation\endlinenomath}

\begin{document}

\title{\bf \boldmath
Search for $e^+e^-\to K^+ K^- \psi(3770)$ at center-of-mass energies from 4.84 to 4.95 GeV
}

\author{
	M.~Ablikim$^{1}$, M.~N.~Achasov$^{4,b}$, P.~Adlarson$^{75}$, O.~Afedulidis$^{3}$, X.~C.~Ai$^{80}$, R.~Aliberti$^{35}$, A.~Amoroso$^{74A,74C}$, Q.~An$^{71,58}$, Y.~Bai$^{57}$, O.~Bakina$^{36}$
% ...
	\\
	\vspace{0.2cm}
	(BESIII Collaboration)\\
	\vspace{0.2cm} {\it
		$^{1}$ Institute of High Energy Physics, Beijing 100049, People's Republic of China\\
% ...
$^{3}$ Bochum Ruhr-University, D-44780 Bochum, Germany\\
		$^{4}$ Budker Institute of Nuclear Physics SB RAS (BINP), Novosibirsk 630090, Russia\\
% ...
$^{35}$ Johannes Gutenberg University of Mainz, Johann-Joachim-Becher-Weg 45, D-55099 Mainz, Germany\\
		$^{36}$ Joint Institute for Nuclear Research, 141980 Dubna, Moscow region, Russia\\
% ...
$^{57}$ Southeast University, Nanjing 211100, People's Republic of China\\
		$^{58}$ State Key Laboratory of Particle Detection and Electronics, Beijing 100049, Hefei 230026, People's Republic of China\\
% ...
$^{71}$ University of Science and Technology of China, Hefei 230026, People's Republic of China\\
% ...
$^{74}$ University of Turin and INFN, (A)University of Turin, I-10125, Turin, Italy; (B)University of Eastern Piedmont, I-15121, Alessandria, Italy; (C)INFN, I-10125, Turin, Italy\\
		$^{75}$ Uppsala University, Box 516, SE-75120 Uppsala, Sweden\\
% ...
$^{80}$ Zhengzhou University, Zhengzhou 450001, People's Republic of China\\
		\vspace{0.2cm}
% ...
$^{b}$ Also at the Novosibirsk State University, Novosibirsk, 630090, Russia\\
% ...
	}
}

\begin{abstract}
Using $e^+e^-$ collision data, corresponding to an integrated luminosity of $892\,\rm pb^{-1}$ collected at center-of-mass energies from 4.84 to 4.95\,GeV with the BESIII detector, we search for the process $e^+e^-\to K^+ K^- \psi(3770)$ by reconstructing two charged kaons and one $D$ meson from $\psi(3770)$. No significant signal of $e^+e^-\to K^+ K^- \psi(3770)$ is found and the upper limits of the Born cross sections are reported at 90\% confidence level.
\end{abstract}

\maketitle


\oddsidemargin  -0.2cm
\evensidemargin -0.2cm

\section{Introduction}

The charmonium states ($c\bar{c}$) provide an excellent experimental laboratory for understanding the non-perturbative mechanism of quantum chromodynamics (QCD).  The conventional charmonium states, like $J/\psi$, $\psi(2S)$, and $\psi(3770)$, have played a significant role in studying the behavior of quarks and gluons. In recent years, more vector states have been observed in the charmonium energy region, which challenges the conventional charmonium states predicted by the quark potential model~\cite{quarkmodel}. Some of them, such as $\psi(4040)$, $\psi(4160)$ and $\psi(4415)$, have been observed by analyzing the line shape of the cross sections of electron-positron annihilating into inclusive hadron states~\cite{BES:2007zwq} that are dominated by open-charm processes; others, such as $Y(4230)$, $Y(4390)$, and $Y(4660)$, have been discovered via hidden-charm final states~\cite{paper2, paper3, BESIII:2021njb,Belle:2008xmh}. 
According to the quark potential model~\cite{Kher:2018}, the masses of the 5S and 6S vector charmonium states are around 4.6~GeV and 5.2~GeV, respectively. In the mass region of $4.7$~GeV to $4.95$~GeV, there should be only one vector charmonium $4^3D_1$ state with mass about 4.8~GeV. For the exotic states, there are different theoretical predictions based on various hypotheses. For example, in Ref.~\cite{Dong:2021juy}, a vector $\Xi_c \bar{\Xi}_c$ molecule is predicted, whereas lattice QCD does not expect such a state~\cite{Chiu:2005ey}. To resolve this controversial situation, further experimental and theoretical investigations are required.

Recently, BESIII reported a new structure around 4.79 GeV in the line shape of the cross sections of the process $e^+ e^- \to D^{*+}_s D^{*-}_s$~\cite{BESIII:2023wsc}. This structure appears to be consistent with the structure observed around 4.71 GeV in BESIII's previous measurement of $e^+ e^-\to K^+K^- J/\psi$~\cite{Y4710} and is also evident in $e^+ e^-\to K_S^0 K_S^0 J/\psi$~\cite{BESIII:2022kcv}, considering the large uncertainties of both mass and width. These new observations inspire us to search for a new process $e^+ e^- \to K^+ K^- \psi(3770)$. The process $e^+ e^- \to \pi^+\pi^-\psi(3770)$ has been studied in the BESIII experiment~\cite{pipi3770,pipi37701,PIPIDD}. Searching for new high-mass vector states involving both $\psi(3770)$ and kaons would provide valuable information to determine the nature of these new structures. Although there is no definite theoretical prediction for $e^+ e^- \to K^+ K^- \psi(3770)$ (Ref.~\cite{Brambilla:2022hhi} only predicts the inclusive width of $Y(4710)$ decays into $J/\psi$ or $\psi(3686)$), a naive presumption suggests that this process would be suppressed in conventional charmonium~\cite{Qian:2023taw} or $[cs][\bar{c}\bar{s}]$ tetraquark assumptions~\cite{Lu:2016cwr}, while an enhancement is expected in the $f_0(980)\psi(2S/1D)$ molecule model~\cite{Guo:2008zg}.

In this paper, for the first time we search for the process $e^+ e^- \to K^+ K^- \psi(3770)$ by using an integrated luminosity of $892\,\rm pb^{-1}$ $e^+e^-$ collision data collected at center-of-mass (CM) energies from 4.84 to 4.95\,GeV with the BESIII detector~\cite{lum_bes3}. The CM energy and corresponding luminosities are listed in Table~\ref{crossection}.

\section{BESIII detector and Monte Carlo}

The BESIII detector~\cite{BESIII} records symmetric $e^+e^-$ collisions 
provided by the BEPCII storage ring~\cite{Yu:IPAC2016-TUYA01}
in the center-of-mass energy range from 2.0 to 4.95~GeV,
with a peak luminosity of $1 \times 10^{33}\;\text{cm}^{-2}\text{s}^{-1}$ 
achieved at $\sqrt{s} = 3.77\;\text{GeV}$. 
The cylindrical core of the BESIII detector covers 93\% of the full solid angle and consists of a helium-based
 multilayer drift chamber~(MDC), a plastic scintillator time-of-flight
system~(TOF), and a CsI(Tl) electromagnetic calorimeter~(EMC),
which are all enclosed in a superconducting solenoidal magnet
providing a 1.0~T magnetic field.
The solenoid is supported by an
octagonal flux-return yoke with resistive plate counter muon
identifier modules interleaved with steel. The
charged-particle momentum resolution at $1~{\rm GeV}$ is
$0.5\%$, and the specific energy loss ($dE/dx$) resolution is $6\%$ for the electrons
from Bhabha scattering. The EMC measures photon energies with a
resolution of $2.5\%$ ($5\%$) at $1$~GeV in the barrel (end cap)
region. The time resolution of the TOF barrel part is 68~ps, while the end cap TOF system is upgraded in 2015 with multi-gap resistive plate chamber technology, providing a time resolution of 60 ps ~\cite{update1,update2}.

Simulated samples produced with a {\sc
	Geant4}-based~\cite{geant4} Monte Carlo (MC) package, which
includes the geometric description~\cite{detvis} of the BESIII detector and the
detector response, are used to determine the detection efficiency
and to estimate backgrounds. The simulation includes the beam
energy spread and initial state radiation (ISR) in the $e^+e^-$
annihilations modeled with the generator {\sc
	kkmc}~\cite{kkmc}.
	The inclusive MC sample includes the production of open charm
processes, the ISR production of vector charmonium(-like) states,
and the continuum processes incorporated in {\sc
kkmc}~\cite{kkmc}. All particle decays are modelled with {\sc
evtgen}~\cite{evtgen} using branching fractions 
either taken from the
Particle Data Group~\cite{PDG2022}, when available,
or otherwise estimated with {\sc lundcharm}~\cite{lundcharm}. Final state radiation~(FSR)
from charged final state particles is incorporated using the {\sc
photos} package~\cite{photos}.

\section{EVENT SELECTION}

The dominant decay channel of $\psi(3770)$ is \mbox{$\psi(3770) \to D\bar D$}.
For the signal process $e^+e^-\to K^+ K^- \psi(3770)$, we only reconstruct two charged kaons and one $\bar{D}$ meson, the presence of $D$ mesons is inferred by
the recoiling mass of $K^+ K^- \bar{D}$. To reconstruct the $\bar{D}$ mesons, we select nine decay modes with large BF and clean backgrounds, including 
three modes $\bar D^0 \to K^+\pi^-$, $\bar D^0 \to K^+\pi^-\pi^0$
and $\bar D^0 \to K^+\pi^+\pi^-\pi^-$ with a summed up to a total 26.6\% branching fraction of $\bar D^0$ decays;
six modes $D^-\to K^+\pi^-\pi^-$, $D^-\to K^{+}\pi^{-}\pi^{-}\pi^0$, $D^-\to K_{S}^0\pi^-$, $D^-\to K_{S}^0\pi^-\pi^0$, $D^-\to K_{S}^0\pi^-\pi^-\pi^+$, and $D^-\to K^+K^-\pi^-$ with a summed up to 27.2\% branching fraction of $D^-$ decays. Throughout the whole text, charge conjugated decays are always implied.

Charged tracks detected in the MDC are required to be within a polar angle ($\theta$) range of $|\rm{cos\theta}|<0.93$, where $\theta$ is defined with respect to the $z$-axis,
which is the symmetry axis of the MDC. For charged tracks not originating from $K_S^0$ decays, the distance of closest approach to the interaction point (IP) must be less than 10\,cm along the $z$-axis, $|V_{z}|$,  and less than 1\,cm in the transverse plane, $|V_{xy}|$.
Particle identification~(PID) for charged tracks combines measurements of the energy deposited in the MDC~(d$E$/d$x$) and the flight time in the TOF to form likelihoods $\mathcal{L}(h)~(h=p,K,\pi)$ for each hadron $h$ hypothesis.
Charged kaons and pions are identified by comparing the likelihoods for the kaon and pion hypotheses, $\mathcal{L}(K)>\mathcal{L}(\pi)$ and $\mathcal{L}(\pi)>\mathcal{L}(K)$, respectively.

Each $K_{S}^0$ candidate is reconstructed from two oppositely charged tracks satisfying $|V_{z}|<$ 20~cm.
The two charged tracks are assigned
as $\pi^+\pi^-$ without imposing further PID criteria. They are constrained to
originate from a common vertex and are required to have an invariant mass
within $(0.486,0.510)~{\rm GeV}/c^2$. This mass window corresponds to about three times the mass resolution. The
decay length of the $K^0_S$ candidate is required to be greater than
twice the vertex resolution away from the IP.


The $\pi^0$ candidates are reconstructed via $\pi^0\to\gamma\gamma$. 
Here, photon candidates are identified using showers in the EMC.  The deposited energy of each shower must be more than 25~MeV in the barrel region ($|\cos \theta|< 0.80$) and more than 50~MeV in the end cap region ($0.86 <|\cos \theta|< 0.92$).  To exclude showers that originate from charged tracks, the angle subtended by the EMC shower and the position of the closest charged track at the EMC
must be greater than 10 degrees as measured from the IP. To suppress electronic noise and showers unrelated to the event, the difference between the EMC time and the event start time is required to be within [0, 700]\,ns.
A one-constraint (1C) kinematic fit is performed to constrain the invariant mass of photon pair to the nominal mass of $\pi^0$~\cite{PDG2022}, and  and a list of $\pi^0$ candidates is prepared for subsequent event selection.

If there are multiple reconstructed $\bar{D}$ candidates, the one with the closest mass to the nominal $\bar{D}$ mass~\cite{PDG2022} is selected. A 1C kinematic fit is then carried out with the $\bar{D}$ meson's nominal mass constraint to improve the resolution of the recoiling mass spectrum and to reduce the background. The $\chi^{2}$ value of the 1C kinematic fit is set to be less than 13.
This selection criterion is optimized by maximizing the Punzi figure of merit, $\frac{S}{\sqrt{B}+\alpha/2}$~\cite{opt}, where $S$ is the number of signal from the signal MC sample, $B$ is the estimated background yield from the inclusive MC samples, and $\alpha$, set at 3, is the expected significance. The $K^\pm$ candidates, which meet the particle identification criteria and possess the lowest momentum among those not utilized in $\bar{D}$ reconstruction, are assumed to be the bachelor kaons unrelated to $D$ or $\bar{D}$ decays.

Even though the process $e^+ e^- \to \phi \psi(3770)$ is highly suppressed in a electron-positron collision experiment due to C-parity violation,
we require the $|M_{K^+K^-}-M_{\phi}|>0.02~{\rm GeV}/c^2$ to reduce the backgrounds including $\phi$ meson in the final states, in which the $M_{K^+K^-}$ and $M_{\phi}$ denote the invariant mass of $K^+K^-$ and $\phi$ nominal mass, respectively. After imposing all event selection criteria, we compare the recoiling mass distributions, $R(K^+K^-\bar D)$ and $R(K^+K^-)$, between data and inclusive MC sample, as shown in Fig~\ref{fig:DD}. Here, $R(K^+K^-\bar D)=\sqrt{(P_{e^+e^-}-P_{K^+}-P_{K^-}-P_{\bar D})^2}$ and $R(K^+K^-)=\sqrt{(P_{e^+e^-}-P_{K^+}-P_{K^-})^2}$ are the recoil
mass of the $K^+K^-\bar D$ and $K^+K^-$, respectively, where $P_{e^+e^-}$, $P_{K^\pm}$ and $P_{\bar D}$  are the 4-momenta 
of the initial $e^+e^-$ system, the $K^\pm$ and $\bar D$, respectively. In general, the data can be well described by the inclusive MC sample.

\section{DATA ANALYSIS}
\begin{figure*}[htbp]
	\centering
    \includegraphics[width=0.99\linewidth]{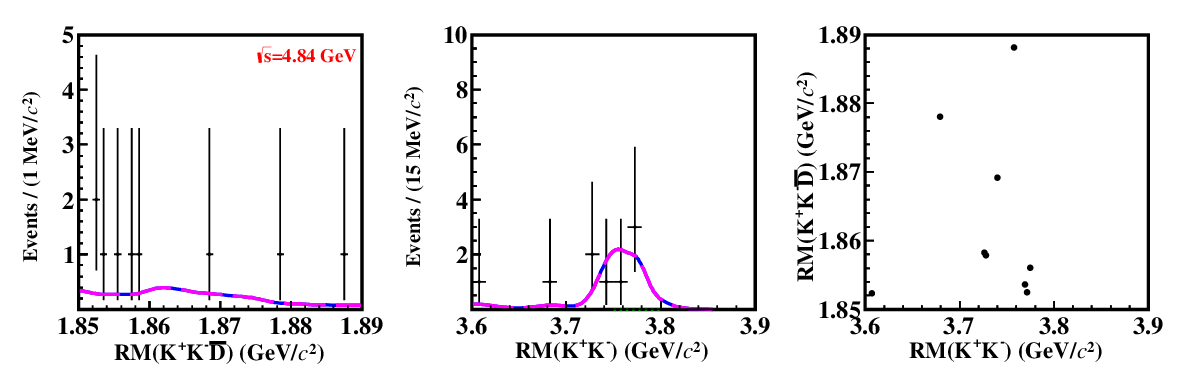}
	\includegraphics[width=0.99\linewidth]{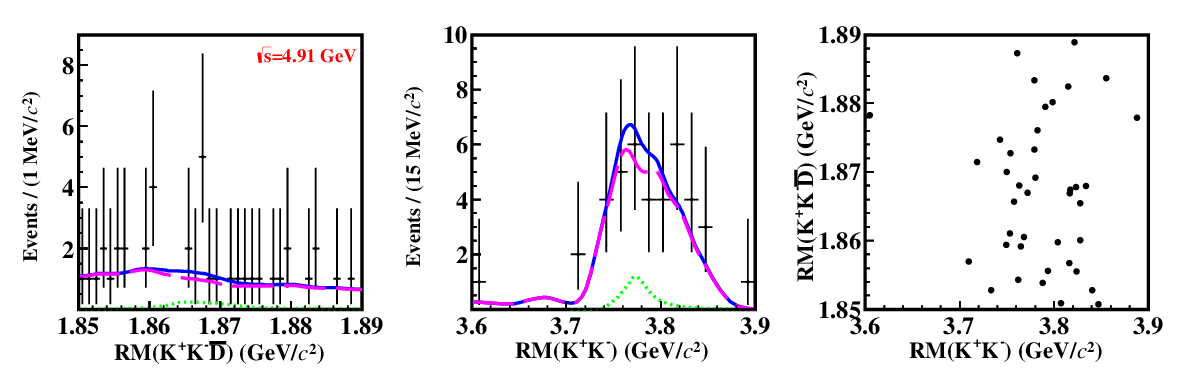}
	\includegraphics[width=0.99\linewidth]{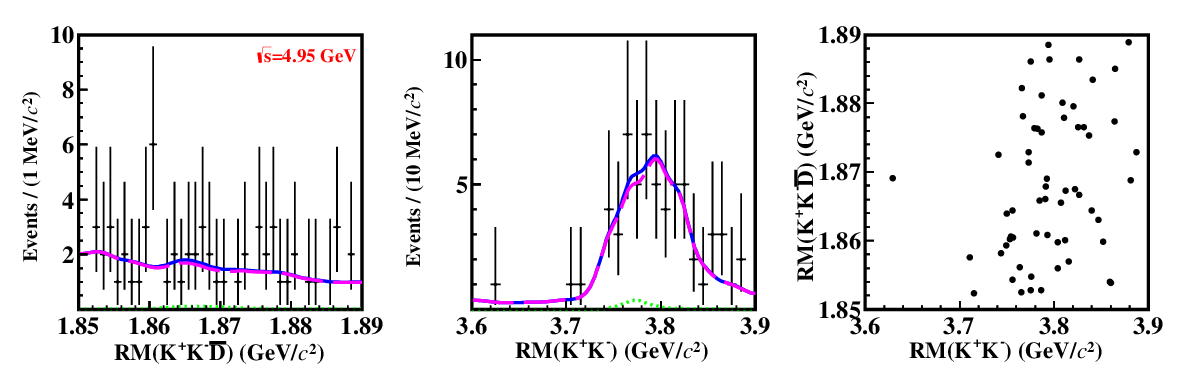}
	\caption{The recoiling mass distributions of $K^+K^-\bar{D}$ (left panel) and $K^+K^-$ (middle panel), and their 2D scatter plot (right panel) at $\sqrt{s}=4.84$, $4.91$, and $4.95$ GeV.
		In the left two columns, the dots with error bars are data, the blue solid lines represent the fitted curve, the green dotted lines represent signal shapes, the pink dashed lines present backgrounds. Notice at $\sqrt{s}=4.84$, the signals are too tiny to be visible.	}
	\label{fig:DD}
\end{figure*}

To extract the signal yields, a two-dimensional (2D) unbinned maximum likelihood fit
is performed to the recoiling mass distributions $RM(K^+K^-\bar{D})$ versus $RM(K^+K^-)$. 
The 2D probability density function (PDF) for the signal is taken from the signal MC simulation. The PDFs of background contributions are extracted from the inclusive MC samples. The dominant backgrounds come from $K^+K^- D\bar D \pi^0 $ and $K^+K^- D\bar D \pi^0\pi^0$ processes. 
Neither unexpected structure nor peaking background is found from the inclusive MC in the fitting region.  To further test the reliability of background shape, we compare the $RM(K^+K^-\bar{D})$ and $RM(K^+K^-)$ distributions (include signal region and non-signal region) between data and MC simulation without any requirement on the $\chi^2$ of $D$ kinematic fit. No significant difference between data and inclusive MC sample is found.
Since no significant signal of $e^+e^- \to K^+K^- D \bar D$ is observed and the $D\bar D$ invariant mass distribution of $e^+e^- \to K^+K^- D \bar D$ is similar with that of $\psi(3770)$ because of the limited phase space, this component is ignored in the fits.
Figure~\ref{fig:DD} shows the $RM(K^+K^-\bar{D})$ versus $RM(K^+K^-)$ fitting results at CM energies $\sqrt{s}=4.84$, $4.91$, and $4.95$ GeV.

In the fits at each energy points, the cross sections of \mbox{$e^+e^-\to K^+ K^- \psi(3770)$} with $\psi(3770) \to D^0 \bar{D}^0$ and $\psi(3770) \to D^+ D^-$ are constrained to be the same. Therefore, after considering the detection efficiency and BF of $\bar{D}$ decays, the ratios of the signal yields in the charged mode relative to the neutral modes are fixed to  $f_{4.84}=0.46$, $f_{4.91}=0.53$, and $f_{4.95}=0.54$ for the three energy points, respectively.

The Born cross section is calculated as,
\begin{equation}
\small
	\sigma^B (e^+ e^- \to K^+ K^- \psi(3770))=\frac{N_{
	\rm sig}}{2\mathcal L_{\rm int}(1+\delta (s))\frac{1}{|1-\Pi|^2}\epsilon_{\rm sig}B_{\rm sub}},
	\label{crossformula}
\end{equation}
where $N_{\rm sig}$ is the number of signal events in the data, $\mathcal L_{\rm int}$ is the integrated luminosity of data measured by Bhabha events~\cite{lum_bes3}, $\epsilon_{\rm sig}$ is the detection efficiency, $B_{\rm sub}$ is aggregate of the BFs of $\psi(3770)\to D\bar D $ and subsequent $D$ decays across the nine tag modes, $\frac{1}{|1-\Pi|^2}$ is the vacuum polarization factor~\cite{EPJC}, and $1+\delta (s)$ is the radiative correction factor defined as
\begin{equation}
	1+\delta (s)=\frac{\int \sigma(s(1-x))F(x,s)dx}{\sigma(s)},
\end{equation}
where $F(x,s)$ is radiator function calculated by QED with accuracy 0.1\%~\cite{QED}. $\sigma$ is the cross section.

\begin{figure}[htpb]
	\centering
	\includegraphics[width=0.9\linewidth]{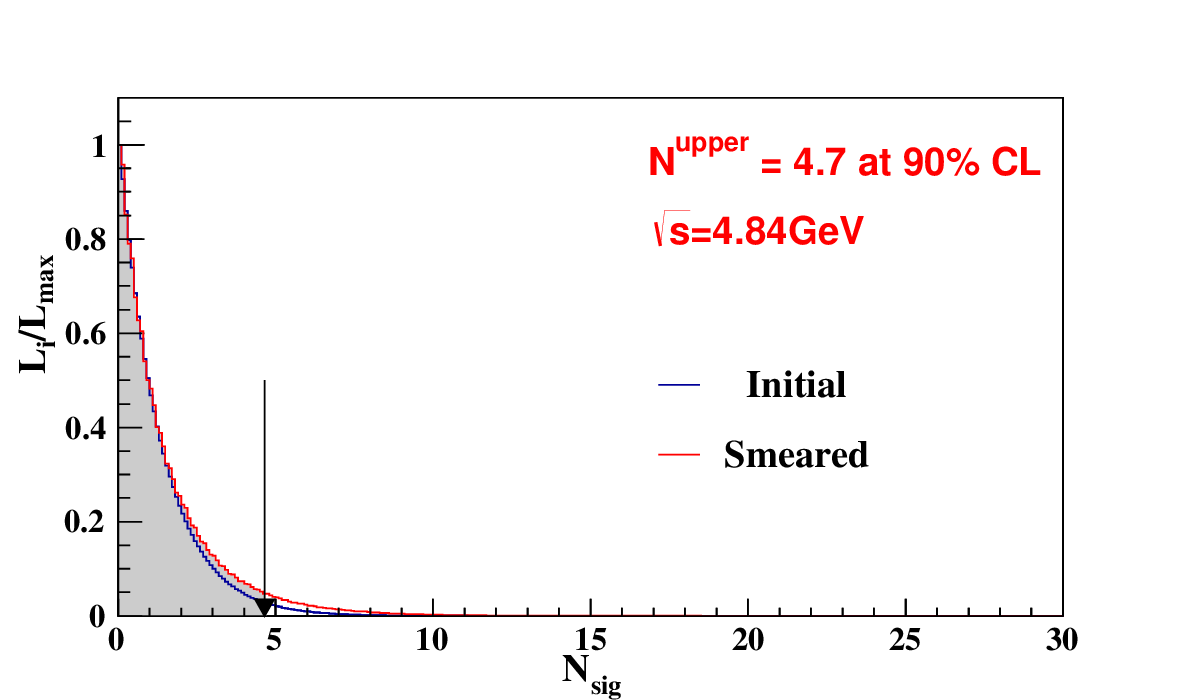}
	\includegraphics[width=0.9\linewidth]{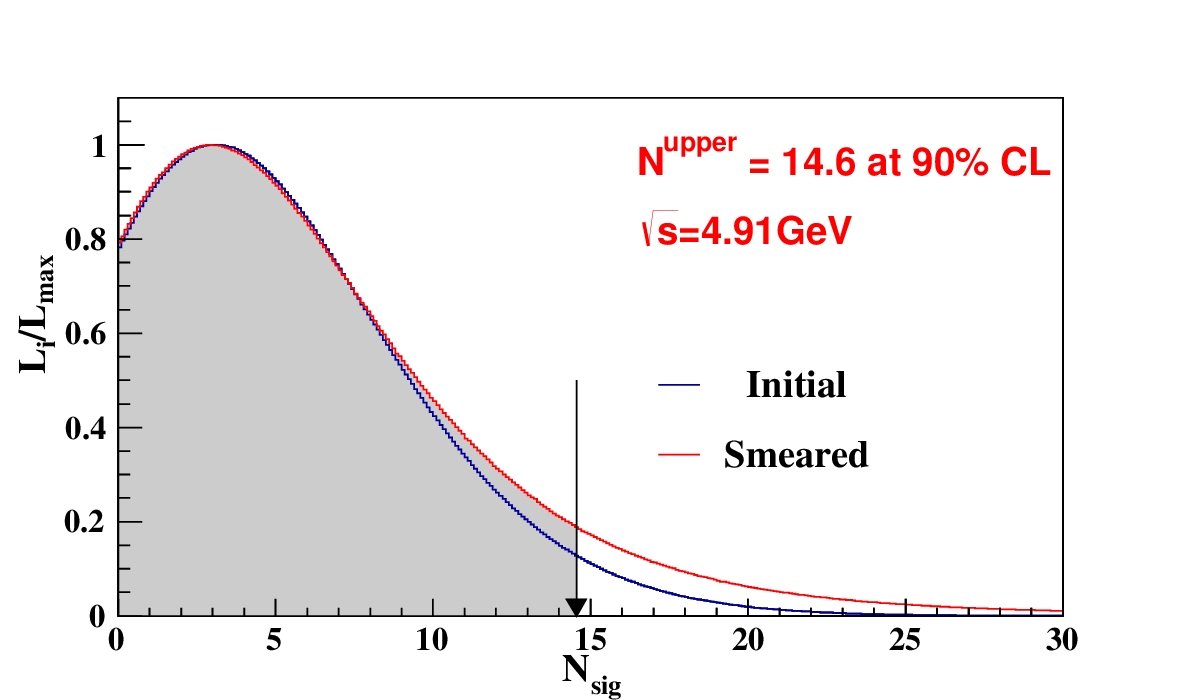}
	\includegraphics[width=0.9\linewidth]{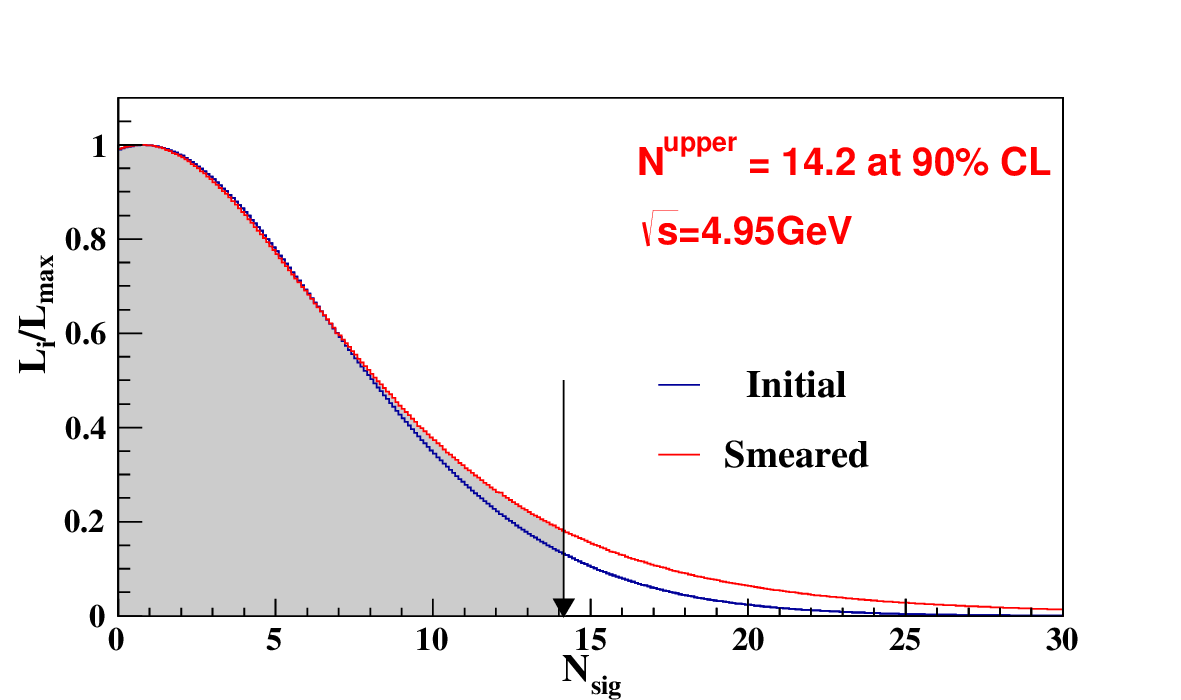}	
	\caption{Normalized likelihood ($L/L_{max}$) as a function of the number of signal events.
		The red and blue solid curves are the likelihood distributions before and after involving the multiplicative systematic uncertainties,
		respectively. The black arrow shows the upper limit of the number of events at 90\% confidence level after considering the systematic uncertainties.}
	\label{fig:upper}
\end{figure}

No obvious signal is found for the process \mbox{$e^+e^-\to K^+ K^- \psi(3770)$}. Utilizing the Bayesian approach~\cite{li} and incorporating the systematic uncertainties, that will be discussed in Sec.~\ref{sec:sys}, we estimate the upper limits on the Born cross sections for $e^+e^-\to K^+ K^- \psi(3770)$ at 90\% confidence level at these three energy points. The corresponding likelihood distributions are shown in Fig.~\ref{fig:upper}, and the upper limit results are listed in Table~\ref{crossection}.

\begin{table*}[htpb]
	\centering
	\caption{Related quantities at each energy point. Here $\sqrt{s}$ is the CM energy, $\mathcal L_{\rm int}$ is the integrated luminosity, $N_{\rm sig}$ is the
		number of signal events by the best fit, $\epsilon_{\rm sig}$ is the efficiency including the BFs of $\psi(3770)$ and $\bar{D}$ decays, $(1+\delta(s))$ is the radiative correction factor, $\frac{1}{|1-\Pi|^2}$ is the vacuum polarization, and $\sigma^{B}$ is the upper limit of the Born cross section, and $N^{\rm upper}$ is the upper limit of the signal yield after considering systematic uncertainties.}
	\begin{tabular}{|c|c|c|c|c|c|c|c|}
		\hline
		$\sqrt{s}~ {\rm (GeV)}$    & $\mathcal L_{\rm int}(\rm {pb}^{-1})$&$N_{\rm sig}$ &$N^{\rm upper}$& $\epsilon_{\rm sig} (\%)$&$(1+\delta(s))$&$\frac{1}{|1-\Pi|^2}$&$\sigma^{B}$(pb) \\   \hline
		4.84   & 525.16$\pm$2.78&$0.00^{+0.56}_{-0.54}$&$<$4.7&0.487$\pm$0.015 &0.727 &1.056&$<$1.2\\
		4.91   & 207.82$\pm$1.10&$3.0^{+4.2}_{-5.2}$&$<$14.6&1.473$\pm$0.027&0.757 &1.056&$<$3.0\\
		4.95   & 159.28$\pm$0.84&$0.8^{+6.1}_{-5.0}$&$<$14.2&1.736$\pm$0.028&0.772 &1.056&$<$3.1\\
		\hline
	\end{tabular}
	\label{crossection}
\end{table*}

\section{SYSTEMATIC UNCRTAINTY}
\label{sec:sys}
In the measurements of the Born cross sections, the systematic uncertainties are categorized into
additive and multiplicative uncertainties. The additive uncertainties originate
from the fit to the $RM(K^+K^-\bar{D})$ versus $RM(K^+K^-)$, that directly affects the fitting results. The multiplicative uncertainties are associated with the efficiencies, and will affect the cross section calculation.

The additive uncertainties originate from the 2D fits, which are primarily influenced by the uncertainty in the signal and background shapes. The uncertainty associated with the signal shape is negligible. In the nominal determination of the upper limits, the background shape is derived from the inclusive MC simulation. To assess the corresponding systematic uncertainty, a 2D polynomial function is utilized as an alternate option. The 2D polynomial function is formed by the product of two polynomial functions corresponding to $RM(K^+K^-\bar{D})$ and $RM(K^+K^-)$. For the $RM(K^+K^-\bar{D})$ dimension, we have experimented with a constant and a 1st-order polynomial function, whereas for the $RM(K^+K^-)$ dimension, we have explored 2nd and 3rd-order polynomial functions. The resulting upper limits based on these two background shapes are detailed in Table~\ref{crossection}, and the larger values are selected for conservation.

The multiplicative uncertainties include luminosity ($\mathcal L_{\rm int}$), $K^\pm$ and $\pi^\pm$ tracking and PID, $\pi^0$ reconstruction,
radiative correction factor ($1+\delta(s)$), $K_S^0$ reconstruction, 1C kinematic fit, MC statistics, and quoted BFs.  The integrated luminosity is measured using Bhabha events, the uncertainty is about 0.6\%~\cite{lum_bes3}. The $K^{\pm}$ tracking and PID efficiencies are estimated by double tag hadronic $D \bar D$ events~\cite{kev}. The data/MC efficiency differences are weighted by the corresponding momentum spectra from signal MC events. The $K^{\pm}$ tracking and PID systematic uncertainties are determined to be 8.5\% and 2.9\%, respectively. The systematic uncertainties of $\pi^{\pm}$ tracking and PID are cited from previous work, the systematic uncertainty is 1.6\% for both tracking and PID~\cite{PI}. In this work, the $\pi^{0}$ selection criteria are the same as those used in
Ref.~\cite{epjc76}. We assign 1.0\% as the systematic uncertainty per $\pi^0$ reconstruction, which includes the effect of photon selection, mass window and 1C kinematic fit. The systematic uncertainty of radiative correction factor is estimated by comparing the difference between factors obtained by the phase space line shape and flat line shape, and is determined to be 10.0\%, 10.5\%, and 11.0\% for $\sqrt{s}=4.84$, $4.91$, and $4.95$ GeV, respectively. The systematic uncertainty of quoted BFs is 10\%, including the $\psi(3770) \to D\bar D$ and $\bar{D}$ subsequent decays~\cite{PDG2022}. The systematic uncertainty of $K_S^0$ reconstruction is 0.2\%, quoting from Ref.~\cite{kev}.
For the uncertainty of kinematic fit, we try to correct the track helix parameters in the MC simulation so as to describe the data better. The correction factors are obtained by using control sample $J/\psi \to \phi f_{0}(980)$~\cite{paper980}. 
The efficiency difference before and after the correction, 1\%, is taken as the systematic uncertainty of the kinematic fit. 
The uncertainties of MC statistics are 3.1\%, 1.8\%, and 1.6\%
for $\sqrt{s}=4.84$, $4.91$, and $4.95$ GeV, respectively. 
The uncertainties associated with the physics model are estimated by comparing the detection efficiencies with those of phase space and $e^+e^-\to f_0(980)\psi(3770)\to K^+K^-\psi(3770)$. The relative differences, which are $15.6\%$, $17.6\%$, and $20.0\%$ for $\sqrt{s}=4.84$, $4.91$, and $4.95$ GeV, respectively, are reported as the corresponding systematic uncertainties.
Adding each systematic uncertainty in quadrature, we obtain the total systematic uncertainties for each energy point. All the multiplicative systematic uncertainties are summarized in Table~\ref{tab:relsysuncertainties}. The blue and red lines in Fig.~\ref{fig:upper} show 
with and without considering the systematic uncertainty likelihood distribution. 

\begin{table}[htp]
	\centering
	\caption{
		Multiplicative systematic uncertainties (\%) in the measurements of the Born cross sections.}
	\label{tab:relsysuncertainties}
	\centering
	\small
	\begin{tabular}{c|c|c|c}
		\hline
			\hline
		Uncertainty&4.84 GeV&4.91 GeV&4.95 GeV\\
		\hline
		Luminosity                         &0.6&0.6&0.6\\
		$K^\pm$ tracking                   &8.5&8.5&8.5\\
		$K^\pm$ PID                        &2.9&2.9&2.9\\
		$\pi^\pm$ tracking                 &1.6&1.6&1.6\\
		$\pi^\pm$ PID                      &1.6&1.6&1.6\\
		Radiative correction factor        &10.0&10.5&11.0\\
		Quoted $\mathcal B$                &10.0&10.0&10.0\\
		$K_{S}^{0}$ reconstruction           &0.2&0.2&0.2\\
		Kinematic fit                       &1.0&1.0&1.0\\
		MC statistics                      &3.1&1.8&1.6\\
		Physics model                      &15.6&17.6&20.0\\
		\hline
		Total                              &23.2&24.7&26.7\\
		\hline
			\hline
	\end{tabular}
\end{table}

\section{SUMMARY}
In this paper, we search for the process $e^+e^- \to K^+K^-\psi(3770)$ using 896 $\rm {pb}^{-1}$ of $e^+e^-$ annihilation data taken at CM energies from 4.84 to 4.95 GeV.
No significant signal for $e^+e^-\to K^+ K^- \psi(3770)$ is observed. Therefore, we determine the upper limits of the Born cross sections at 90\% confidence level to be $1\sim 3$ pb.
We also notice that the upper limits reported in this paper are obviously smaller than the cross sections measured in the process $e^+e^-\to \pi^+ \pi^- D^+ D^-$~\cite{PIPIDD}, which could be attributed to a combination of factors, including the suppressed phase space and strange quark production in $e^+ e^- \to K^+ K^- \psi(3770)$. But we don't know which factor is dominant at present. Further measurements based on additional data samples within and beyond this energy region, that will be collected at BESIII in the future, may help to clarify this matter.

\section{ACKNOWLEDGMENTS}

The BESIII Collaboration thanks the staff of BEPCII and the IHEP computing center for their strong support. This work is supported in part by National Key R\&D Program of China under Contracts Nos. 2020YFA0406300, 2020YFA0406400; National Natural Science Foundation of China (NSFC) under Contracts Nos. 11635010, 11735014, 11835012, 11935015, 11935016, 11935018, 11961141012, 12025502, 12035009, 12035013, 12061131003, 12192260, 12192261, 12192262, 12192263, 12192264, 12192265, 12221005, 12225509, 12235017; the Chinese Academy of Sciences (CAS) Large-Scale Scientific Facility Program; the CAS Center for Excellence in Particle Physics (CCEPP); Joint Large-Scale Scientific Facility Funds of the NSFC and CAS under Contract No. U1832207; CAS Key Research Program of Frontier Sciences under Contracts Nos. QYZDJ-SSW-SLH003, QYZDJ-SSW-SLH040; 100 Talents Program of CAS; The Institute of Nuclear and Particle Physics (INPAC) and Shanghai Key Laboratory for Particle Physics and Cosmology; European Union's Horizon 2020 research and innovation programme under Marie Sklodowska-Curie grant agreement under Contract No. 894790; German Research Foundation DFG under Contracts Nos. 455635585, Collaborative Research Center CRC 1044, FOR5327, GRK 2149; Istituto Nazionale di Fisica Nucleare, Italy; Ministry of Development of Turkey under Contract No. DPT2006K-120470; National Research Foundation of Korea under Contract No. NRF-2022R1A2C1092335; National Science and Technology fund of Mongolia; National Science Research and Innovation Fund (NSRF) via the Program Management Unit for Human Resources \& Institutional Development, Research and Innovation of Thailand under Contract No. B16F640076; Polish National Science Centre under Contract No. 2019/35/O/ST2/02907; The Swedish Research Council; U. S. Department of Energy under Contract No. DE-FG02-05ER41374.

\end{document}